\documentclass[aps,pra,twocolumn,amssymb,amsfonts,showpacs,groupedaddress,floatfix,nofootinbib]{revtex4}

\usepackage{graphicx}
\usepackage{epsfig}
\usepackage[english]{babel}

\newcommand{\ignore}[1]{}
\newcommand{\mComment}[1]{}
\newcommand{\lComment}[1]{}
\renewcommand{\mComment}[1]{\textcolor{blue}{Manny: #1}}
\renewcommand{\lComment}[1]{\textcolor{red}{Lorenza: #1}}

\def\cc{{\cal C}}
\def\cd{{\cal D}}
\def\ce{{\cal E}}
\def\cl{{\cal L}}

\def\ch{{\cal H}}
\def\cs{{\cal S}}

\def\cq{{\cal Q}}

\def\ca{{\cal A}}
\def\cr{{\cal R}}

\def\cv{{\cal V}}
\def\cy{{\cal Y}}
\def\cz{{\cal Z}}
\def\ur{\smash{\upharpoonright}}
\newcommand{\cmplxs}{{\mathbb C}}

\begin{document}

\title{ A note on verification procedures for quantum noiseless subsystems } 

\author{ Lorenza Viola }
\email{ lviola@lanl.gov } 
\author{Emanuel Knill}
\email{knill@lanl.gov}
\affiliation{ Los Alamos National Laboratory, Mail Stop B256, 
Los Alamos, New Mexico 87545 }

\date{\today}

\begin{abstract}

We establish conditions under which the experimental verification of 
quantum error-correcting behavior against a linear set of error operators 
$\ce$ suffices for the verification of noiseless subsystems of
an error algebra $\ca$ contained in $\ce$. From a practical standpoint, 
our results imply that the verification of a noiseless subsystem need not 
require the explicit verification of noiseless behavior for all possible 
initial states of the syndrome subsystem.
\end{abstract}


\maketitle

\section{Introduction}

Noiseless subsystems (NSs) provide a comprehensive conceptual framework
for understanding stabilization strategies for quantum information 
\cite{knill:qc2000a,viola:qc2000b,zanardi:qc2001,Zanardi-Lloyd}. NSs 
include as a special case decoherence-free subspaces (DFSs) 
\cite{DuanGuo,Zanardi-NoiselessCodes,LidarDFS}. Because, physically, 
the occurrence of NSs requires the presence of symmetry in the underlying 
noise process, NSs may not exist for arbitrary error models. However, 
if the appropriate symmetry requirements are met, the protection that 
NSs can afford is especially powerful, as encoded information remains 
immune to errors indefinitely in time. In the language of general quantum 
error correction (QEC) \cite{knill:qc2000a}, the latter property 
characterizes NSs as infinite-distance quantum error-correcting codes. 

An important issue in both analyzing error-correcting performance and
implementing error-control benchmarks
\cite{Knill-Benchmark,Knill-5BitQEC} is to establish operational
criteria under which NS-behavior may be reliably diagnosed from
available data. While it might seem that the implementation of a
desired NS simply amounts to verifying that information is preserved
once appropriately encoded, two considerations make the procedure less
straightforward in practice. On one hand, due to unavoidable
operational imperfections, the implemented decoding transformation may
differ from the intended one in unknown ways, making the actual
subsystem identification potentially inequivalent to the abstract
noiseless one. On the other hand, a proper NS is paired with a
non-trivial syndrome subsystem in such a way that a distinct
infinite-distance quantum error-correcting code can be associated with
every initial state of the syndrome. Accordingly, the verification of
an NS appears at first to require that noiselessness of the relevant
information is checked for all possible syndrome initializations.  

It is the purpose of this work to address these issues from an
experimentally motivated perspective and to point out a criterion that
is applicable whenever stability of encoded information under an error
algebra is experimentally verified. The content of the paper is organized 
as follows. In Sect. II we recall the subsystem view of QEC, by emphasizing 
in particular the difference between finite- and infinite-distance 
error-correcting behavior and the association of the latter with noiseless 
degrees of freedom. Sect. III is devoted to the formulation of the 
verification problem and to the construction of verification procedures 
within the assumed setting. This is done by first discussing how 
error-correcting behavior against a generic linear set of errors may be 
diagnosed in a typical QEC experiment, and then by showing how stronger 
conclusions may be reached if an algebra is contained in the error set. 
In Sect. IV, the general analysis and results are illustrated by revisiting 
the prototypical example of a three-qubit NS for collective noise as introduced 
in \cite{knill:qc2000a} and implemented in \cite{Viola:qc2001a,FortunatoNS}. 
The example also demonstrates how the algebraic structure may be exploited for 
deducing the action of errors when the syndrome subsystem is initialized to 
states other than the one explicitly implemented or, equivalently, for 
inferring the stability of the relevant information under a larger error
set than explicitly checked -- both features being advantageous from the 
practical point of view. The paper concludes with a brief summary in Sect. V.

\section{Error-correcting subsystems} 

A general description of QEC, applicable 
to both finite- and infinite-distance error control, is offered by the
subsystem approach \cite{Knill97a,knill:qc2000a,viola:qubit,LAScience1}.
Let $S$ be a finite-dimensional system, with state space $\cs$, dim($\cs) = d$, 
and let $\ca_S=\text{End}(\cs)$ denote the corresponding operator algebra. We
may assume $\cs \simeq \cmplxs^d$, and $\ca_S \simeq \text{Mat}_d (\cmplxs)$. 
For example, $d=2^n$ for an $n$-qubit system. Note that while we will make 
explicit reference to the usual qubit setting in the present discussion, 
more general situations ({\it e.g.}, involving higher-dimensional subsystems) 
could easily be handled. Suppose that $S$ is used to protect $k$ qubits, 
$k < \log_2(d)$, against a known family of error operators $\ce =\{ E_a \}$. 
We require that the ``no-error'' event is correctable, thus $\ce$ contains the 
identity. Because quantum correctability by a given error control strategy is 
preserved under linear transformations \cite{Knill97a}, we also assume that the 
error space $\ce$ is a linearly closed subset of operators in $\ca_S$. The 
subsystems view of QEC relies on separating the degrees of freedom representing 
the logical state from the ones encoding the effect of the errors on the intended 
code $\cc \subset \cs$.  Thus, $\ce$ is correctable by $\cc$ provided that an 
isomorphism exists, 
\begin{equation}
\omega :\, \cs \rightarrow \cl \otimes \cz  \oplus \cd \:,
\label{subsys0}
\end{equation}
such that for every $E\in \ce$ and every $|\psi\rangle_\cc \in \cc$, 
\begin{equation}
E |\psi\rangle_\cc = \omega^{-1} ( |\psi\rangle_\cl \otimes 
|\varphi_E \rangle_\cz) \:,
\label{error}
\end{equation}
for a vector $|\varphi_E \rangle_\cz \in \cz$ only dependent upon $E$. 
Because $\openone$ is correctable, the code associated to $\cl$
is the subspace 
\begin{equation}
\cc = \omega^{-1}( \cl \otimes |\varphi_0\rangle_\cz )\:,
\label{code}
\end{equation}
with $|\varphi_0\rangle_\cz $ corresponding to no error. Note that $\cc$ 
can detect errors which cause leakage into $\cd$, but not correct
them. Thus, under the condition that all errors in $\ce$ are correctable, 
the mapping (\ref{subsys0}) effectively singles out a subspace 
$\ch = \omega^{-1}( \cl \otimes \cz) \subseteq \cs$. The
correspondence between states in $\ch$ and states in 
$\cl \otimes \cz$ under the restriction $\omega \ur \ch$ {\sl defines} the 
subsystem identification of the QEC procedure, $\cl$ and $\cz$ being the 
information-carrying and syndrome subsystem, respectively. Given any 
correctable error $E \in \ce$, it is then possible to describe the action
of $E$ directly in the subsystem representation by introducing an operator 
$\tilde{E} = \omega \circ E \circ \omega^{-1}$. Equivalently, one may write 
\begin{equation}
E |\Psi \rangle_\ch = \omega^{-1} 
\Big(\tilde{E} \big(|\psi \rangle_\cl \otimes |\varphi\rangle_\cz\big) \Big) \:,
\label{tildeX}
\end{equation}
with $\omega |\Psi \rangle_\ch = |\psi \rangle_\cl \otimes |\varphi\rangle_\cz$.
We shall use a similar notation for operator sets {\it e.g.}, 
$\tilde{\ce}=\{ \tilde{E}\, |\, E \in \ce \}$, {\it etc.}

Combining Eqs. (\ref{error})-(\ref{code}), correctability of $\ce$ by $\cc$ 
is equivalent to the condition that errors in $\ce$ affect only the syndrome
subsystem when the latter is appropriately initialized. Note, however, that the 
assumed linear structure of $\ce$ does {\sl not} suffice in general to 
guarantee that cumulative errors from $\ce$ remain correctable, unless 
information is properly returned to $\cc$ by resetting $\cz$ to its reference 
state $|\varphi_0\rangle_\cz$. While active recovery is a necessary feature of 
finite-distance QEC, codes with stronger error-correcting properties may be 
designed if $\ce$ is known to have additional structure. Suppose that 
$\ce \supseteq \ca$, where $\ca$ is a $\dagger$-closed sub-algebra of $\ca_S$ 
containing $\openone$. Because $\ca$ is closed under operator multiplication, 
{\sl arbitrary} cumulative errors remain in $\ca$. Under these conditions, 
$\cc$ becomes an infinite-distance quantum error-correcting code for $\ca$ 
\cite{knill:qc2000a}. 

In fact, infinite-distance behavior is associated with the existence
of protected degrees of freedom supported by NSs of $S$
\cite{knill:qc2000a}. Within the error-algebraic framework, a state
space decomposition of the form (\ref{subsys0}) emerges through the
reduction of $\ca$ into irreducible components
\cite{knill:qc2000a,zanardi:qc2001}.  Accordingly, $\ch$ can be
identified with a fixed invariant subspace, $\cd$ with its orthogonal
complement, and the noiseless factor $\cl$ carries an irreducible
representation of the commutant $\ca'$ of $\ca$. Whenever information
is protected using a NS, resetting of the syndrome $\cz$ becomes
unnecessary, hence no active intervention is required for maintaining
information \cite{Comment}. In the simplest instance, which is
realized by a DFS, this happens because the syndrome state effectively
does {\sl not} evolve under the errors, thus the relevant syndrome
subsystem is one-dimensional, immediately identifying $\cl \simeq \cc$
as an infinite-distance quantum error-correcting code.  In a generic
NS case, where both the logical and the syndrome factors $\cl$ and
$\cz$ are non-trivial subsystems, information encoded in $\cl$ is
protected irrespective of the evolution experienced by $\cz$. This
implies that a proper NS is associated with a distinct
infinite-distance quantum code for every reference state of the
latter.

\section{Verification setting} 

In practice, taking advantage of the ability of a given code to
protect information against errors in $\ce$ requires implementing a
decoding procedure capable of restoring the information contained in
the code after errors in $\ce$ occur. Let $\cq$ be the state space of
the physical subsystem $Q$ of $S$ that carries the quantum information
after decoding.  To be specific, if $S$ consists of $n$ qubits, dim
$\cq$ = dim $\cl$ = $2^k$, and we can treat the remaining $n-k$ qubits
as ancillae, with an associated state space $\ch_a=(\cmplxs^2)^{\otimes (n-k)}$.  
The QEC procedure is then implemented by first
appropriately initializing the ancillae state, next by transferring
the resulting $k$-dimensional input space to the intended code $\cc$
through an encoding operation, and then, after an error event happens,
using a decoding operation to extract the state of $\cq$ \cite{LAScience1}.

Let $U_{d}, U_{e}$ denote the experimentally implemented decoding and
encoding operations, respectively. While the decoding is designed so as to 
provide a realization of the abstract mapping $\omega$ given in Eq. (\ref{subsys0}), 
in practice, operational errors and inaccuracies will prevent one from exactly 
knowing the actual $U_d$. Yet, by construction, $U_d$ provides a 
subsystem identification of the form 
\begin{equation}
U_d :\, \cs \rightarrow \cq \otimes \cy  \oplus \cr \:,
\label{subsys1}
\end{equation}
where $\cy$ is the state space of a physical syndrome degree of freedom carrying 
the effect of the errors, and $\cr$ collects the states of $\cs$ for which 
the extraction of the relevant information by $U_d$ effectively
fails. Eventually, one would like to claim that (\ref{subsys1}) realizes an 
error-correcting subsystem equivalent to (\ref{subsys0}), at least in the case 
where a unique subsystem with the correct behavior is known to exist for $\ce$
except for irrelevant unitary rotations in the underlying factors. 
But how can we actually verify that the subsystem identified by the implemented
decoding is noiseless under the error model of interest, and under what conditions 
can we conclude that a desired NS has been realized?

Let us consider a verification setting defined by the following assumptions:
\begin{itemize}
\item[ 1)] The ancillae are prepared in a known pure state, say $|0\rangle_a 
\in \ch_a$.
\item[ 2)] The error model $\ce$ is known.
\item[ 3)] The implemented decoding transformation, $U_d$, is unitary.
\item[ 4)] The initial state in the code $\cc$ is recovered perfectly for all 
$E \in \ce$.
\end{itemize}
While the first requirement is always necessary for the implemented QEC procedure 
to be meaningful, the remaining conditions may or may not enter the definition
of the setting in principle. None of the requirements can be rigorously fulfilled 
in actual implementations. Apart from the assumed perfect fidelity in both the 
initialization and the recovery steps, one could naturally encounter 
situations where either 2) or 3) (or both) would need to be relaxed to some 
extent. We focus here on the simplest verification scenario, having in mind 
device technologies capable to meet all the requirements 1) through 4) with 
sufficiently high accuracy. In particular, the present analysis is directly 
motivated by the recent experimental implementations of DFSs 
\cite{Kwiat-DFS,Kielpinski-DFS,Fortunato-DFS} and NSs 
\cite{Viola:qc2001a,FortunatoNS} using single-photon optics, trapped
ions, and liquid-state NMR.

\subsection{Verification for finite-distance codes} 

Let $\ce$ be a generic linear error set and suppose that one has 
experimentally verified that information protected by the implemented code 
$\cc$ is recovered perfectly for an error basis 
$\{ E_\ell \} = \{E_0=\openone, E_1, \ldots \}$ of $\ce$ {\it i.e.}, 
we have observed that 
\begin{equation}
U_d \big( E_\ell |\psi\rangle_\cc \big) =
|\psi\rangle_\cq \otimes |\varphi_{E_\ell}\rangle_\cy \:, 
\hspace{2mm} \forall \ell \:, 
\label{decode}
\end{equation}
for arbitrary encoded states $|\psi\rangle_\cc$. Then stability under all error 
operators in $\ce$ can be immediately inferred using linearity. Note that the 
existence of a non-trivial summand $\cr$ in the correspondence (\ref{subsys1})
is signaled by the fact that the span$\{ |\varphi_{E_\ell}\rangle_\cy \}$ does 
not cover the full ancilla state space $\ch_a$. In the assumed qubit setting, 
this means that the syndrome qubits may be in general a proper subset of the 
ancillary ones.  
From Eq. (\ref{decode}), one knows that, in particular, 
the $\openone$ is correctable, namely 
\begin{equation}
U_d \big( |\psi\rangle_\cc \big) =
|\psi\rangle_\cq \otimes |\varphi_r \rangle_\cy \:, 
\label{decode0}
\end{equation}
for some reference state $|\varphi_r \rangle_\cy \in \cy$. 
This makes two remarks possible: $(i)$ since the prepared state 
$|0\rangle_a$ of the ancillae is pure, and $U_d$ is unitary, one can check 
to what extent the implemented $U_e$ is unitary by verifying the purity of 
the decoded state in Eq. (\ref{decode0}). Suppose that based on this  
observation we can take $U_e$ to be unitary with high accuracy henceforth.
$(ii)$ In the identification provided by $U_d$, the encoding operation $U_e$ 
implies the initialization of the syndrome subsystem in the state 
$|\varphi_r\rangle_\cy$, and the code $\cc$ may be represented as
\begin{equation} 
\cc = U_d^{-1} ( \cq \otimes |\varphi_r\rangle_\cy )\:.
\label{expcode}
\end{equation}
If desired, the accuracy to which $U_e$ avoids transferring unintended
information to $\cr$ may be checked by measuring the 
amplitude in states orthogonal to the span $\{U_d^{-1} (|\psi\rangle_\cq 
\otimes |\varphi_{E_\ell}\rangle_\cy ) \}$. Finally, under the identification 
given by $U_d$, Eqs. (\ref{decode}) and (\ref{expcode}) together imply that 
errors in $\ce$ have an identity action on the logical subsystem when $\cy$ is 
initialized to $|\varphi_r \rangle_\cy$ {\it i.e.}, one can conclude that for 
all $E \in \ce$
\begin{equation}
\tilde{E} \big(|\psi \rangle_\cq \otimes |\varphi_r \rangle_\cy \big)
= |\psi \rangle_\cq \otimes |\hat{E}(\varphi_r) \rangle_\cy
\:, \;\;\forall |\psi \rangle_\cq 
\in \cq \:,
\label{idAct}
\end{equation}
or, equivalently, 
\begin{equation}
U_d (E \ur {\cc}) =\tilde{E} \ur {\cq \otimes |\varphi_r\rangle_\cy}
= \openone_\cq \otimes \hat{E} \:, 
\label{idAct2}
\end{equation}
with $\ur$ denoting restriction, $\tilde{E}= U_d E U_d^{-1}$, and 
$ \hat{E} |\varphi_r\rangle_\cy =  | \hat{E}(\varphi_r) \rangle_\cy$. 
This means that experimentally establishing Eq. (\ref{idAct}) or (\ref{idAct2})
suffices for claiming that $\cc$ is a quantum $\ce$-correcting code. 
Whenever a unique $k$-dimensional error-correcting code is known to
exist for given $n$ and $\ce$, the implemented code is effectively the one 
abstractly described by Eqs. (\ref{subsys0})-(\ref{error}).

\subsection{Verification for infinite-distance codes} 

In addition to having 
verified the validity of Eqs. (\ref{idAct})-(\ref{idAct2}), assume now the
stronger condition that the errors include a known, non-trivial error algebra
$\ca$, with $\ce \supseteq \ca$ and $\openone \in \ca$. Then the 
algebraic structure of $\ca$ further enables one to infer a 
trivial action of errors on the logical subsystem when the syndrome subsystem is 
initialized to certain states other than $|\varphi_r \rangle_\cy$. Let 
\begin{equation}
\cv = \{\, | \hat{E}(\varphi_r) \rangle_\cy \:|\: E \in \ca \, \} \subseteq \cy
\end{equation}
denote the states of $\cy$ reachable from $|\varphi_r \rangle_\cy$ under 
the effect of error operators in $\ca$.  Thus, $\cv$ depends upon 
$|\varphi_r \rangle_\cy$ and $\ca$. The fact that $\cv$ is a linear space 
follows from the property that, for $E_1,E_2 \in \ca$ and 
for arbitrary complex $\alpha, \beta$, one may write 
\begin{equation}
\alpha |\hat{E}_1 (\varphi_r)\rangle_\cy +  
\beta  |\hat{E}_2 (\varphi_r)\rangle_\cy =
|\hat{E}( \varphi_r)\rangle_\cy \:, 
\end{equation}
with ${E}= \alpha {E_1} + \beta  {E_2} \in \ca$. 
We can then prove the following 

\vspace{2mm}

{\bf Theorem:} Let $\ce \supseteq \ca$, $\ca$ being an error algebra 
on $S$, and let $\cv$ be defined as above. Assume that stability under
$\ce$ has been verified as in Eq. (\ref{idAct2}). Then
\begin{equation} 
U_d \Big(\ca \ur { U_d^{-1}(\cq \otimes \cv ) } \Big) = 
\tilde{\ca} \ur {\cq \otimes \cv} \subseteq 
\openone_\cq \otimes \text{End}(\cv) \:. 
\label{main}
\end{equation} 

Proof: We need to show that any error operator in $\ca$ has no effect on 
$\cq$ whenever the state of $\cy$ is in $\cv$. Let $|\chi\rangle_\cv
\in \cv$. Then $|\chi\rangle_\cv= \hat{E}_b |\varphi_r \rangle_\cy$ 
for some $E_b \in \ca$. If $E_a$ and $|\psi \rangle_\cq$ are 
any error operator in $\ca$ and state in $\cq$, respectively, one has: 
$\tilde{E}_a |\psi \rangle_\cq \otimes |\chi \rangle_\cv = 
\tilde{E}_a |\psi \rangle_\cq \otimes \hat{E}_b |\varphi_r \rangle_\cy
= \tilde{E}_a \tilde{E}_b |\psi \rangle_\cq \otimes |\varphi_r \rangle_\cy 
= \tilde{E}_{ab} |\psi \rangle_\cq \otimes |\varphi_r \rangle_\cy =
|\psi \rangle_\cq \otimes |\hat{E}_{ab}(\varphi_r) \rangle_\cq$,
for some $\tilde{E}_{ab}= \tilde{E}_a \tilde{E}_b \in \ca$.
\hspace*{\fill}\mbox{\rule[0pt]{1.4ex}{1.4ex}}

\vspace{2mm}

According to the above Theorem, $\cv$ effectively determines the portion
of the syndrome's state space $\cy$ relative to which noiselessness of 
$\cq$ against $\ca$ may be inferred from a verification procedure based on 
a {\sl fixed reference state} $|\varphi_r\rangle_\cy$ or, equivalently, a 
{\sl fixed encoding} $U_e$. Because the error model is assumed to be known, 
the dimensionality of $\cv$ may be inferred from the observed behavior of 
the syndrome subsystem upon decoding. Three different possibilities may
arise:
\begin{itemize}

\item $1= \text{dim}(\cv) < \text{dim}(\cy)$. This implies that 
$\cv =\text{span}\{ |\varphi\rangle_\cv\} $ for a fixed state in $\cv$ 
independent (up to an irrelevant phase factor) of the error operator in 
$\ca$. Because $\ca$ contains the $\openone$, then $|\varphi\rangle_\cv = 
|\varphi_r \rangle_\cv$, meaning that the state of the syndrome 
subsystem is {\sl invariant} under $\ca$. Having verified Eq. (\ref{idAct2}), 
one knows that $\cc$ realizes an infinite-distance error-correcting code 
for $\ca$. With $\cq \otimes \cv \simeq \cq$ and 
$\text{End}(\cv) \simeq \cmplxs$, Eq. (\ref{main}) reads
\begin{equation} 
U_d \Big(\ca \ur {U_d^{-1}(\cq) }\Big) = 
\tilde{\ca} \ur {\cq } \subseteq \openone_\cq \:,
\end{equation} 
which is exactly the characterization of a DFS against $\ca$ 
\cite{Zanardi-NoiselessCodes,lidarDFSQEC,kempe:qc2001a}. Thus, observing
that information is robustly encoded against $\ce \supseteq \ca$, and 
that $\ca$ preserves the syndrome's state, implies the verification of 
$\cq$ as an infinite-distance DFS-code for $\ca$. Note that establishing 
$\cq$ as a (proper) NS under $\ca$ would require verifying DFS-behavior for 
a set of linearly-independent reference states spanning $\cy$.

\item $1< \text{dim}(\cv) < \text{dim}(\cy)$. In this case, the above 
Theorem implies that verifying a trivial action of errors in $\ca$ on $\cc$ 
according to Eqs. (\ref{idAct})-(\ref{idAct2}) suffices for inferring a trivial 
action of $\ca$ whenever the reference state of $\cy$ is an arbitrary state 
in $\cv$. Thus, one can conclude that any quantum code $U_d^{-1}(\cq \otimes 
|\chi\rangle_\cy)$, $|\chi\rangle_\cy \in \cv \subset \cy$, provides 
infinite-distance error protection against $\ca$. In a sense, the
procedure establishes $\cq$ as a (proper) NS against $\ca$ 
{\sl conditionally} on initialization of $\cy$ in $\cv$.  
 
\item $1< \text{dim}(\cv) = \text{dim}(\cy)$. Because $\cv \subseteq \cy$, 
one has $\cv =\cy$, implying that every state in $\cy$ is effectively 
reachable from $|\varphi_r \rangle_\cy$ through the action of an error in 
$\ca$. Under these circumstances, the verification of stability under $\ce$ 
implies via the Theorem that noiselessness can be inferred {\sl irrespective} 
of the state of $\cy$. If a unique $k$-dimensional NS with dim $\cy$ = dim $\cz$ 
is known to exist, the procedure enables one to conclude that $\cq$ is 
effectively the intended NS against $\ca$. 
\end{itemize}

\section{Example} 

Let us briefly illustrate the above ideas on the simplest 
instance of a non-trivial quantum NS, which arises when a system of three 
qubits is used to protect a qubit in the presence of arbitrary collective 
noise \cite{knill:qc2000a,viola:qubit}. 
In this case, $\cs \simeq \cmplxs^8$, $\ca_S \simeq \text{Mat}_8(\cmplxs)$, 
$k=1$, and the relevant subsystem decomposition (\ref{subsys0}) applies 
to the subspace $\ch_{1/2} $ of states carrying total spin angular 
momentum $J^2 =j(j+1)$, $j=1/2$. $\cl$ and $\cz$ are both two-dimensional, 
with $\cl=\text{span}\{ |\ell\rangle_\cl\,|\, \ell=0,1\}$ and 
$\cz=\text{span}\{ |j_z\rangle_\cz\,|\, j_z=\pm1/2 \}$, $\ell$ and 
$j_z$ being a logical quantum number and the total $\hat{z}$-angular momentum 
eigenvalue, respectively. The summand $\cd = \ch^\perp_{1/2}= 
\ch_{3/2}$ is the four-dimensional subspace of states symmetric under qubit 
exchange, corresponding to $j=3/2$. Explicit realizations of the correspondence 
$\omega:\, \cs \rightarrow \cl \otimes \cz \oplus \ch_{3/2}$ are given in 
\cite{viola:qubit,LAScience1,FortunatoNS,kempe:qc2001a}. 

$\cl$ is designed as a NS against the collective error algebra $\ca_c$, 
which contains all possible permutationally-invariant error operators. 
For three qubits, $\ca_c$ is a twenty-dimensional non-abelian sub-algebra
of $\ca_S$, supporting $\cl$ as a unique NS (up to unitary transformations). 
Practically relevant abelian sub-algebras of $\ca_c$ include 
$\ca_x,\ca_y, \ca_z$, describing collective error processes about a fixed 
spatial axis. Each of the latter sub-algebras is linearly spanned by the set 
of four Hermitian Kraus operators describing a full-strength phase damping 
channel $\ce_u$ along the direction $u$ {\it e.g.}, 
$\ca_z =\text{span}\{ K_a^z\,|\, a=0,\ldots3\}$, 
$\varrho_{out}= \ce_z(\varrho_{in}) = \sum_{a=0}^3 K_a^z \varrho_{in}
K_a^z$, and so forth. While complete expressions for $K_a^u$, 
$a=0,\ldots,3, u=x,y,z$ may be found in \cite{FortunatoNS}, the 
representation in the collective error-correcting subsystem
decomposition is especially transparent. For collective $z$ errors, 
for instance, one obtains that $\tilde{K}_0^z =\tilde{K}_1^z=0$, and 
$${\tilde{K}_2^z} \ur {\ch_{1/2}}  =  
\openone_\cl \otimes |+1/2\rangle \langle +1/2 |_\cz \:,$$
\vspace*{-8mm}
\begin{equation}
{ \tilde{K}_3^z} \ur {\ch_{1/2}}  =  
\openone_\cl \otimes |-1/2\rangle \langle -1/2 |_\cz \:,
\end{equation}
corresponding to full-strength phase damping on the syndrome subsystem 
$\cz$ alone. Similar representations hold for $u=x,y$ \cite{FortunatoNS}. 
Let us also denote by $\ce_{vu}$ a composite error process obtained by 
cascading $\ce_u$ and $\ce_v$ in sequence \cite{FortunatoNS}. A set of 
operation elements for such a composite process can be constructed by 
multiplication of the sets describing the individual error components. 

In experimental realizations of the above NS as in 
\cite{Viola:qc2001a,FortunatoNS}, the implemented decoding $U_d$
effectively maps the abstract $\cl$, $\cz$ degrees of freedom to a
physical information-carrying qubit $\cq$ and a physical syndrome
qubit $\cy$, respectively. In the resulting identification, the initialization 
of the syndrome subsystem $\cy$ is typically constrained to a fixed state 
$|\varphi_r\rangle_\cy $ determined by the implemented encoding.
In the setting of \cite{Viola:qc2001a,FortunatoNS}, the presence of 
unintentional amplitude in the $\ch_{3/2}$ subspace is reflected
in the state of the non-syndrome ancilla qubit upon decoding. 

Various verification procedures for the intended NS may be considered
depending on the experimentally available class of error processes. 
Suppose, for instance, that we have verified Eq. (\ref{idAct2}) under 
arbitrary single-axis collective errors, namely under the error set 
\begin{equation}
\ce=\text{span}\{K_a^x, K_b^y, K_c^z \,|\, a,b,c=0,\ldots,3 \} \:, 
\end{equation}
in terms of the above-mentioned collective Kraus operators. Suppose, in 
addition, that by looking at the behavior of the decoded syndrome qubit 
one is able to determine that dim$(\cv_x)=\text{dim}(\cv_y)=2$, 
whereas dim$(\cv_z)=1$. This effectively implies initialization of the 
system in a $j_z$-eigenstate, say $|\varphi_r\rangle_\cy = 
|+ 1/2 \rangle_\cy$.  
Thus, by using the Theorem, one can conclude that 
the decoded signal originates from a proper NS under $\ca_x$ and $\ca_y$, 
and from a DFS under $\ca_z$. However, by the same argument used in the 
proof of the Theorem, the fact that stability under the two error processes 
$\ce_x,\ce_y$ has been verified irrespective of the initial syndrome state
implies the possibility to {\sl enlarge the relevant error set} to include
arbitrary products of $x,y$ error operators. This effectively 
enables one to deduce the validity of the condition (\ref{idAct2}) under 
a linear set $\ce' \supseteq \ce$ larger than the one explicitly tested
{\it i.e.}, 
\begin{equation}
\ce' = \text{span}\{K_a^x K_b^y, K_b^y K_a^x,
K_a^x K_c^z, K_b^y K_c^z \}\:,
\end{equation}
where errors of the form $K_c^z K_a^x, K_c^z K_b^y$ are absent because 
stability under $\ce_z$ can only be assumed conditionally on the initial 
invariant state $|+1/2\rangle_\cy$. 
Finally, one can show that $\ce' \supseteq  \ca_c$, hence by applying again 
the Theorem it is possible in fact to infer noiselessness of the implemented 
subsystem $\cq$ against the full $\ca_c$. 

A second, more direct, verification procedure consists of checking stability 
of the encoded information under two composite, conjugate error processes 
$\ce_{vu}$ and $\ce_{uv}$, $\ce_{uv} = \ce_{vu}^\dagger$, and by using the 
fact that the resulting error set, 
\begin{equation}
\ce'' = \text{span}\{K_a^u K_b^v, K_b^v K_a^u \}\:,
\end{equation}
again satisfies the property that $\ce{''} \supseteq  \ca_c$. 
Finally, if DFS-behavior with initialization into the orthogonal state 
$|\varphi_r\rangle_\cy = |-1/2\rangle_\cy$ is observed as well, then 
verification of robust 
behavior under $\ca_x, \ca_y,\ca_z$ again directly translates into 
verification of the desired NS-behavior against $\ca_c$ via the Theorem.

\section{Conclusion} 

We have outlined verification procedures for quantum NSs in a simple 
experimentally motivated setting. As a main practical implication of our 
analysis, establishing a NS need not require the complete verification 
of the initial syndrome space provided that sufficient access to the final 
decoded states is available. This may be practically advantageous to avoid the 
need of checking different encodings for the same error model. Verification 
procedures designed under the assumptions of unitary decoding and known error 
behavior, as well as perfect fidelity as invoked throughout here, may be expected 
to remain valid if the relevant conditions can be met with sufficiently high 
accuracy. However, it is not {\it a priori} obvious that procedures that are 
equivalent (as in the above NS Example) in such an idealized scenario will 
remain applicable and equally reliable when some of the assumptions are 
relaxed {\it e.g.}, implementation is not perfect. In general, identifying and 
characterizing verification procedures for quantum NSs and error-correcting 
codes under realistic constraints is an interesting issue which deserves 
further investigation.

\section{Acknowledgments} 

This work was supported by the DOE (contract W-7405-ENG-36) and by the NSA. 
L.V. gratefully acknowledges the support of the Los Alamos Office of 
the Director through a J.R. Oppenheimer Fellowship.


\begin{thebibliography}{20}
\bibitem[{Com()}]{Comment}
\bibinfo{note}{We assume that the internal evolution of the system can be set
  to zero or included among the primary error generators.}
\end{thebibliography}

\end{document}